# Probing Lattice Dynamics and Electronic Resonances in Hexagonal Ge and Si$_x$Ge$_{1-x}$ Alloys in Nanowires by Raman Spectroscopy

*Diego de Matteis[1], Marta De Luca[1], Elham M. T. Fadaly[2], Marcel A. Verheijen[2], Miquel López-Suárez[3], Riccardo Rurali[3], Erik P.A.M. Bakkers[2], Ilaria Zardo[1]*

[1] Departement Physik, Universität Basel, 4056 Basel, Switzerland

[2] Department of Applied Physics, Eindhoven University of Technology, 5612AP Eindhoven, The Netherlands.

[3] Institut de Ciència de Materials de Barcelona (ICMAB–CSIC), Campus de Bellaterra, 08193 Bellaterra, Barcelona, Spain

ABSTRACT. Recent advances in nanowire synthesis have enabled the realization of crystal phases that in bulk are attainable only under extreme conditions, *i.e.* high temperature and/or high pressure. For group IV semiconductors this means access to hexagonal-phase Si$_x$Ge$_{1-x}$ nanostructures (with a 2H type of symmetry), which are predicted to have a direct band gap for $x$ up to 0.5 - 0.6 and would allow the realization of easily processable optoelectronic devices. Exploiting the quasi-perfect lattice matching between GaAs and Ge, we synthesized hexagonal phase GaAs-Ge and GaAs-Si$_x$Ge$_{1-x}$ core-shell nanowires with $x$ up to 0.59. By combining position-

, polarization- and excitation wavelength-dependent μ-Raman spectroscopy studies with first-principles calculations, we explore the full lattice dynamics of these materials. In particular, by obtaining frequency-composition calibration curves for the phonon modes, investigating the dependence of the phononic modes on the position along the nanowire, and exploiting resonant Raman conditions to unveil the coupling between lattice vibrations and electronic transitions, we lay the grounds for a deep understanding of the phononic properties of 2H-$Si_xGe_{1-x}$ nanostructured alloys and of their relationship with crystal quality, chemical composition, and electronic band structure.

KEYWORDS: Raman spectroscopy, nanowires, hexagonal (lonsdaleite) SiGe, resonant Raman, phonons, crystal structure transfer.

The exploitation of unique optoelectronic, mechanical, and thermal properties of nanostructured materials has a long history, even antecedent its deep understanding and systematic investigation. Indeed, the high surface-to-volume ratio of nanostructures, together with the possibility of quantum confinement, results in interesting behaviors for optoelectronic, thermoelectric, biomedical and many more applications.[1-3] This diversity and abundance also come from the possibility of obtaining materials in crystal structures that are not easy to achieve in bulk forms. For example, GaAs nanowires (NWs) can also be synthesized in the wurtzite (WZ) crystal structure, which can be obtained in bulk only under extreme pressures,[4-6] while the growth mechanisms typical of NWs have provided a powerful route to obtain it. The availability of unusual crystal structures can be relevant even for fields in which there is already a general consensus on the optimal materials for applications. This is the case, for example, for group IV semiconductors. Their use in electronics dates back to the discovery of the *pn* junction and has been so successful that we can now define the invention of the Metal-Oxide-Semiconductor Field-Effect Transistor (MOSFET) as the start of the "silicon revolution".[7] From 1960 to 2018 an estimated $13*10^{21}$ MOSFETs have been fabricated[8] and silicon has undoubtedly gained a central role in integrated circuit technologies and informatics. Nevertheless, Si and Ge hardly breached the field of optoelectronics, as in their bulk crystalline form they exhibit an indirect band gap, resulting in poor light emission properties. On the contrary, as supported by both theoretical and experimental investigations, $Si_xGe_{1-x}$ compounds exhibit direct band gaps (for x up to 0.5 - 0.6) when synthetized in the hexagonal form (with a 2H type of symmetry, also known as lonsdaleite), which can be the starting point for an efficient class of optoelectronic devices.[9,10]

The lack of information on the vibrational properties of these materials and on their dependence on composition calls for the exploration of their lattice dynamics, as the vibrational spectrum is closely related to the chemical composition, the crystal quality, the alloy formation process, and it can also be a probe of the electronic band structure. In this work we present an investigation of such properties performed by means of μ-Raman spectroscopy on single NWs and theoretical calculations based on density functional theory (DFT). Hexagonal core-shell GaAs-$Si_xGe_{1-x}$ NWs with $0 \leq x \leq 0.59$ were synthetized exploiting the quasi-perfect lattice matching between WZ GaAs and lonsdaleite Ge (referred to as 2H in the following) with the crystal structure transfer method.[11,12]

We were able to uncover the composition dependence of the NWs' phononic spectrum, to analyze the homogeneity of the NWs with spatially-resolved measurements, and to probe the electronic band structure of the alloys with resonant Raman scattering experiments. As a consequence, we could obtain a full picture of all the phonon modes of 2H-Ge and $Si_xGe_{1-x}$, frequency-composition calibration curves for the main phonon modes of the $Si_xGe_{1-x}$ alloys, an assessment of the quality and reliability of the NW synthesis process, and finally some insight into the optically active electronic bands' energy and symmetry, as well as into their coupling with lattice vibrations.

RESULTS AND DISCUSSION

**Hexagonal Ge and $Si_xGe_{1-x}$ samples**

We employ the crystal structure transfer method to fabricate 2H-Ge and $Si_xGe_{1-x}$ crystals as reported by Fadaly *et al.*, where the hexagonal crystal structure is adopted from a hexagonal material template.[12] Here, we utilize single crystalline WZ GaAs NWs as a hexagonal template to epitaxially grow high-quality 2H-Ge and $Si_xGe_{1-x}$ crystals in a core-shell geometry. The choice of GaAs NWs as a core is due to the fact that there is almost no lattice mismatch with 2H-Ge and a small one with Ge-rich $Si_xGe_{1-x}$ alloys. To further avoid the induced strain due to high Si content in $Si_xGe_{1-x}$, we use very thin GaAs NWs cores (35-45nm). The studied NW core-shell structures were fabricated in a Metal Organic Vapor Phase Epitaxy (MOVPE). The GaAs NWs were synthesized *via* catalyst assisted growth following the Vapor Liquid Solid (VLS) mechanism using gold (Au) catalyst seeds and from Trimethyl Gallium (TMGa) and Arsine ($AsH_3$) gas precursors. The resulting GaAs NWs are of very high crystal quality with only few stacking faults (on average 2-3 stacking faults/µm, as shown in the extended data of the figures in ref.13). Prior to the $Si_xGe_{1-x}$ shell growth, the Au catalyst seeds were chemically etched to avoid Au contamination in the shell. Then, the $Si_xGe_{1-x}$ shells were grown epitaxially around GaAs NW cores using germane ($GeH_4$) and disilane ($Si_2H_6$) gas precursors. For further details on growth and structural characterization see the Methods section. In the same ref. 13, which shows NWs grown with the nearly the same conditions as the ones studied in this work, atom probe tomography measurements reveal the presence of As in the shell at a concentration of about $10^{19}$ atoms/cm$^3$.

In this study, a wide range of 2H-$Si_xGe_{1-x}$ compositions ($0 < x < 0.59$) were studied; for details of NW dimensions and compositions, see Table S1 and Figure S1 in the Supporting Information 1. Figure 1 shows the structural characterization for two typical samples of the studied composition range ($Si_{0.27}Ge_{0.73}$ and $Si_{0.59}Ge_{0.41}$); pure 2H-Ge was extensively described in ref. 13. Figure 1 (A, B) display two representative Scanning Electron Microscopy (SEM) images of $Si_{0.27}Ge_{0.73}$ and $Si_{0.59}Ge_{0.41}$, respectively. The SEM images of the NW arrays show very small variations in the morphology implying that they are uniform and the SEM images reflect the average morphology across the samples and, accordingly, the structures measured with Raman spectroscopy. To confirm the purely hexagonal crystal structure, a representative high resolution High Angular Annular Dark Field (HAADF) Scanning Transmission Electron Microscopy (STEM) image for GaAs/$Si_{0.59}Ge_{0.41}$ is shown in Figure 1C. The HAADF-STEM image displays an excellent epitaxial relationship between the GaAs and $Si_{0.59}Ge_{0.41}$ and confirms the ABAB stacking of the 2H crystal structure in the [0001] growth direction. To investigate the composition uniformity across the radial direction of the core-shell structure, few NWs from the NW ensembles shown in Figure 1 (A, B) were cut open and thin lamellas were extracted and analyzed using Energy Dispersive X-ray (EDX) spectroscopy as depicted in Figure 1 (D-E). The EDX elemental maps in Fig1. (D-E) and their corresponding line scans in Figure 1F confirm the uniform material composition after radial shell growth resulting in a nominal Si composition of $27 \pm 2\%$ and $59 \pm 4\%$, for these two samples, respectively. We also observe very thin Ge-rich alloy regions along a sunburst-like geometry, aligned to the corners between the NW facets; a known effect in alloyed non-planar materials due to growth kinetic effects.[14-16] We also notice a Ge-rich region around the core, an effect that we are investigating with further experiments and theoretical modelling. The Si percentage in these sunburst regions have nominal compositions of $24.1 \pm 0.6\%$ %, and $58 \pm 1\%$ for the $Si_{0.27}Ge_{0.73}$ and $Si_{0.59}Ge_{0.41}$ samples, respectively, which is within the alloy fluctuation percentage within the core-shell structure (see Figure S2 in the Supplementary Information, **SI1**).

**Composition dependence of the Raman spectra**

We have computed the Raman spectra within density-functional perturbation theory (DFPT) of $Si_xGe_{1-x}$ with $x = 0.0625$, 0.25, and 0.5 in a $2 \times 2 \times 1$ supercell of the 4-atom 2H primitive cell. More details are given in the computational methods section. In the case of x = 0.25 and x = 0.5 we have generated 5-6 independent random configurations corresponding to the target nominal composition and averaged the computed Raman spectra (x = 0.0625 corresponds to one single Si atom in a Ge matrix and therefore averaging is not necessary). In this way, we partially palliate the limitations deriving from a description of a disordered system by using a computational cell that is rather small and can thus be treated efficiently at the first-principles level. Despite these limitations, the theoretical results are a useful guide for the interpretation of the experimental measurements discussed below. Polarization-resolved Raman scattering calculations and experiments were performed in the same backscattering geometry. Considering a {XYZ} reference system where the single NW is aligned to the Z axis, the incident photon wavevector is antiparallel to the X axis and the scattered photon wavevector is parallel to it (see inset of Figure 2). The excitation and detection light polarization vectors, $\varepsilon_i$ and $\varepsilon_s$, lie in the YZ plane, which is the plane of the sample. The scattering configuration is expressed in terms of excitation and detection polarizations in brackets. The calculations are performed either with the exciting and scattered light both polarized along (ZZ) the NW growth axis or with both perpendicular (YY) to it.

The results from DFPT are summarized in Figure 2, together with the calculated spectra of pure 2H-Ge for reference. The spectra are generated by using Lorentzian curves with a full width at half maximum (FWHM) of 3 cm$^{-1}$ to approximately reproduce the experimental FWHM. In the 2H-Ge, in ZZ configuration only one mode is allowed: the longitudinal optical mode, with symmetry $A_{1g}$, at 297.7 cm$^{-1}$. In the YY configuration, there are three modes allowed: an intense transverse optical mode with $E_{2g}$ symmetry at 286.1 cm$^{-1}$, a low-intense transverse optical $E_{1g}$ mode at 297.4 cm$^{-1}$ and the $A_{1g}$ mode at 297.7 cm$^{-1}$ also visible in ZZ configuration. Since $E_{1g}$ and $A_{1g}$ are almost degenerate, they appear as a single peak, and we will label this mode as $A_{1g}/E_{1g}$ mode from now on (for simplicity, we keep this notation also in the ZZ configuration). All these features enable the distinction between group-IV hexagonal and cubic crystals (in the latter no $E_{2g}$ mode exists, and the transverse and longitudinal optical modes are degenerate and have an $F_{2g}$ symmetry, see the calculated Raman spectra of cubic Ge in Figure S3[17]). Let us now focus on the 2H-Si$_x$Ge$_{1-x}$ alloys. As it can be seen in Figure 2, when one Si atom is added ($x$= 0.0625) a spectral feature close to 400 cm$^{-1}$ appears, a clear signature of a vibrational mode associated to the Si-Ge bond. The intensity and frequency of this mode increase with the Si content, tending to the computed values (indicated with two arrows above the top panel) of an artificial 2H-SiGe lattice, *i.e.* a fully ordered crystal system where each Si atom is bonded to four Ge atoms and vice versa, like in GaAs or SiC. When the Si concentration increases further, we also observe the appearance of higher frequency Raman active modes between 450 and 500 cm$^{-1}$, which come from Si-Si bonds and tend to the values of pure 2H-Si (also in this case the Raman active modes of 2H-Si are indicated with two arrows). Notice that these spectral features were absent at $x = 0.0625$ where one single Si atom was present and thus no Si-Si bond could be formed. Averaging over more random configurations or using a much larger computational cell would make the leading Ge-Ge, Si-Ge, and Si-Si peaks stand out more clearly with respect to the secondary peaks, particularly at $x = 0.5$, as indeed happens in the experimental measurements where much larger volumes are probed.

We have then performed Raman scattering experiments to be compared to the calculations. We have measured NWs transferred by using a micromanipulator from the growth substrate to a SiC substrate. Measurements were performed at room temperature with the 514 nm excitation wavelength. More details are given in the experimental methods section. In Figure 3A and 3B we display representative Raman spectra collected on single GaAs-Si$_x$Ge$_{1-x}$ ($x$= 0, 0.13, 0.21, 0.33, 0.42, 0.59, from bottom to top) core-shell NWs in the ZZ and YY geometries. Similarly to the theoretical spectra, the experimental spectra of the alloy NWs are characterized by the presence of three intense features attributed to the Ge-Ge, Si-Ge and Si-Si vibrations in order of increasing Raman shift. These peaks are broad and asymmetric because they result from the convolution with secondary features that we assign to vibrations of one of the aforementioned atom pairs in various chemical environments. These appear as broad and generally weak features, as they result from the averaged contributions of various atomic arrangements[18-20] that locally correspond to deviations from the global stoichiometry. In addition to these effects we also considered the possible influence of the As dopants on the Raman spectra, in terms of phonon frequency shifts and asymmetric broadening of phonon modes (*e.g.* Fano lineshapes). In cubic Si and Ge, levels of As doping similar to ours typically induce a shift of the phonon modes $< 1$ cm$^{-1}$,[21] and we can expect a similar effect on hexagonal Ge and SiGe. Regarding the issue of asymmetric lineshapes, in cubic $Si_xGe_{1-x}$ alloys, it was shown that, as alloy disorder increases (x approaches 0.5), more contributions from phonons at the zone edge appear, which add low energy tails to the Raman peaks. This effect is stronger for the Si-Ge vibrations, just like in Fig. 3A and 3B, where one can see that the Si-Ge band is the most asymmetric.[22] Undoped cubic $Si_xGe_{1-x}$ alloys investigated in refs. 23 and 24 also display lineshapes very similar to ours. All this considered, we attribute the asymmetric appearance of the Raman bands to alloy-induced disorder and convolution between features arising from slightly different chemical environments, as described earlier.

The experimental spectra measured on 2H-Ge are very similar to the calculated spectra displayed in Figure 2. In the YY scattering configuration we can see the transverse optical $E_{2g}$ mode at 287.7 cm$^{-1}$, with a FWHM of 4 cm$^{-1}$, and the degenerate transverse and longitudinal $A_{1g}/E_{1g}$ modes at 299.4 cm$^{-1}$ (FWHM of 3 cm$^{-1}$), with the first component being much more intense than the second. In the ZZ configuration, instead, the $A_{1g}/E_{1g}$ modes are the only ones allowed, and they have the same frequency and FWHM that they have in the YY configuration. We stress that the absence of a lineshape broadening on the phonon modes of 2H-Ge allows us to rule out any major effect of As doping in the determination of the frequency of the phonon modes. Due to the high signal-to-noise ratio of the measurements and the high structural quality of the samples, these values can be considered as reference values for 2H-Ge. Compared to the computed spectra, the small differences in the relative intensities of the peaks are due to experimental effects that are not taken into account in the calculations (*i.e.*, size effects, dielectric mismatch effect, and effect of the finite numerical aperture of the microscope objective[25]).

We performed a quantitative analysis of the spectra as deconvolution of multiple Lorentzian curves, which enables us to find the minimum number of spectral contributions needed to describe the experimental line shape. This analysis allows us to discuss the behaviour of the main bands as well as the assignment of the various features, dividing the frequency range according to the three atom pairs (an example of the performed deconvolution is displayed in Figures S4 and S5 in **SI2**). For discussing the assignment of the peaks, besides those ascribed to the above discussed Ge-Ge, Si-Ge and Si-Si modes, we adopt the notation presented in ref 19, with the reminder that it was developed for a 1D computational description of the possible configurations, while we use it for the general 3D environment. Therefore, we label the peaks as $(A-B)^{A;B}$, where between brackets we state the atoms involved in the vibrating bond, and as superscript we indicate the prevalent element(s) surrounding them. For simplicity, we do not use the subscript indicating sensitivity to first or second neighbours. The necessity of such a labelling stems from the fact that, in alloys, each species can be in different environments. For example, any Si atom can have from zero to four Ge atoms as first neighbours. The five arrangements might not be all distinguishable in the Raman measurements and most importantly in the deconvolution: in those cases, we will employ multiple superscripts, separated by semicolons. The superscripts "SiGe" indicate that the contribution to the spectrum comes from an atom oscillating in a mixed environment.

In the analysis of the spectra, we identify the higher energy components of the bands (more pronounced in ZZ scattering configuration) as the $A_{1g}/E_{1g}$ modes, and the lower energy components (more pronounced in YY scattering configuration) as the $E_{2g}$ modes (see **SI2**). As a result of this analysis, in Figure 3C we show the frequency of the prevalent component of the main bands, according to the different polarization configurations, as a function of the alloy composition together with their fitting curves. The composition dependent frequencies, as derived from fitting the experimental data, and thus strictly valid only for the measured concentration range, are the following:

$A_{1g}/E_{1g}$ modes (in ZZ configuration):

$$\omega_{Si-Si}\,(cm^{-1}) = 461.95 + 50.01\,x \qquad \text{(Eq. 1.1)}$$

$$\omega_{Si-Ge}\,(cm^{-1}) = 392.12 + 48.48\,x - 42\,x^2 \qquad \text{(Eq. 1.2)}$$

$$\omega_{Ge-Ge}\,(cm^{-1}) = 299.03 - 16.67\,x \qquad \text{(Eq. 1.3)}$$

$E_{2g}$ modes (in YY configuration):

$$\omega_{Si-Si}(cm^{-1}) = 461.27 + 45.22\,x \qquad \text{(Eq. 1.4)}$$

$$\omega_{Si-Ge}\,(cm^{-1}) = 388.36 + 34.26\,x - 29\,x^2 \qquad \text{(Eq. 1.5)}$$

$$\omega_{Ge-Ge}\,(cm^{-1}) = 289.37 - 12.56\,x \qquad \text{(Eq. 1.6)}$$

The Si-Si band shows the strongest dependency on the composition and goes linearly towards ~514 cm[-1], the phonon frequency of pure nanostructured 2H Si.[11] The Si-Ge frequencies are the only ones to show a non-linear dependence on the composition. Given the general similarity of our trends to the ones reported in the literature[23,24] for analogous cubic alloys, in which the frequency downshift is reported in the $x > 0.6$ range, we suggest that a quadratic fit is representative, despite the second order coefficient being small for this concentration range. The Ge-Ge Raman shifts decrease linearly starting from the pure 2H-Ge phonon frequency of 299 cm$^{-1}$. The trend of the last three points ($x = 0.33, 0.42, 0.59$) might suggest a damped dependence, as reported by Herman and Magnotta[24] and by Grein and Cardona,[26] but also linear trends have been reported.[23,27]

In agreement with Ref. 28, the different composition dependencies of the three modes can be explained by the competition between two effects. One is the effect deriving from the fact that for any $x$ other than the extremes, the alloy lattice parameters are forced to be larger than pure Si, and smaller than pure Ge. This leads to an upshift of the frequencies as $x$ increases (the lattice is compressed and the modes harden). The other effect is the disorder-induced shorter order range that results in frequency downshifts with respect to an ideal lattice. See Ref. 29for more details; also notice that disorder for Ge-Ge modes increases with $x$, while for Si-Si modes it decreases, because the system approaches pure Si. The strong composition dependence of the Si-Si Raman shift can come from the two effects acting in the same direction. For Ge-Ge vibrations we can imagine the two effects acting against each other, but overall resulting in a slower redshift. Finally, the Si-Ge vibrational frequencies behave similarly (increasing) to Si-Si for $x < 0.5$ and to Ge-Ge (decreasing) for $x > 0.5$, exhibiting thus a non-monotonic trend. Overall, we find a good qualitative agreement with Raman shift *vs* composition trends reported in literature,[19,23,24,27] and we ascribe quantitative discrepancies to the limited concentration range at our disposal. Finally, we stress that the Equations (1.1-1.6) and the consequent discussion would be almost the same in the case of undoped Ge and SiGe, because As dopants are known to introduce a very small shift in the phonon frequencies (< 1 cm$^{-1}$, as discussed above).

In order to discuss the assignation of the spectral features to bond + environment units, we divide the Raman shift range in three sections, according to the three possible atom pairs.

In the highest energy range, *i.e.* 460 cm$^{-1}$ < ω < 490 cm$^{-1}$, we find the Si-Si band. It is composed of two peaks that we can assign to (Si-Si)$^{SiGe;Si}$, *i.e.* main peak at higher frequency, and (Si-Si)$^{Ge}$, *i.e.* lower intensity and lower frequency (Figure S5). Indeed, the frequency of Si-Si vibrations decreases with the increase of Ge concentration (see the discussion above), *i.e.* the weaker peak comes from the oscillation of Si atoms in Ge rich environments while the main peak is the sum of the contributions from Si atoms with zero to two Ge neighbors. The (Si-Si)$^{SiGe;Si}$ peak emerges for $x = 0.21$, though at this composition it has lower intensity than the (Si-Si)$^{Ge}$ peak. These considerations are consistent with the observations of Pages *et al.* in ref19.

The Si-Ge vibrations appear in the intermediate range of 370 cm$^{-1}$ < ω < 450 cm$^{-1}$ (Figure S5). The results of the deconvolution of the spectra are consistent with the picture of four distinguishable environments around a Si-Ge bond. We propose a division of the Si-Ge contributions in three branches and, analogously to what mentioned before for the Si-Si vibrations, we can assign the branches to (Si-Ge)$^{Ge}$, (Si-Ge)$^{SiGe}$, and (Si-Ge)$^{Si}$ environments in order of increasing Raman shift. Regarding (Si-Ge)$^{Si}$, we observe its splitting (highlighted by a circle in Figure 3A), as noted in refs. 19 and 26 for $x \lesssim 0.75$. This results in a "pseudo-doublet" (as it contains far more than two spectral features) and reveals the fourth distinguishable environment. We assign the lower energy component of the pseudo-doublet to Ge-rich environments and the higher energy one to Si-rich environments at the second neighbor scale, whereas we do not discriminate (Si-Ge)$^{Ge}$ and (Si-Ge)$^{Si}$ according to second neighbors. This is the only case in which we have to invoke sensitivity to second nearest neighbors to explain the wealth of peaks. It is also worth noting that we observe another exchange of spectral weight, *i.e.* the equalization of (Si-

Ge)$^{Ge}$ and (Si-Ge)$^{SiGe}$ at $x \sim 0.16$, but we had no means to investigate the one between (Si-Ge)$^{SiGe}$ and (Si-Ge)$^{Si}$ hypothesized for $x \sim 0.84$. Nevertheless, we can state that the lower frequency component of the pseudo-doublet, which at high Si content corresponds to the only component of (Si-Ge)$^{Si}$ (the splitting takes place for $x < 0.75$), grows with the increase in Si content. This is consistent with the possibility that it will eventually become more intense than the main (Si-Ge)$^{SiGe}$ feature.

The Ge-Ge vibrations appear in the 260 cm$^{-1}$ < ω < 300 cm$^{-1}$ (Figure S4). In addition to the main intense feature, the deconvolution reveals the presence of a second, weaker, one. With arguments similar to those used for Si-Si modes and derived from the analysis of Figure 3C, we attribute the weak components observed between 260 and 280 cm$^{-1}$ (depending on the value of $x$ and on the polarization) to (Ge-Ge)$^{Si}$, which leaves the main Ge-Ge feature to be (Ge-Ge)$^{SiGe;Ge}$. The analysis of the lower energy component is, however, complicated by the rise of a very broad, possibly disorder-induced, band between 150 and 260-270 cm$^{-1}$. The E$_{2g}$ and A$_{1g}$/E$_{1g}$ modes can only be distinguished by the deconvolution analysis for $x > 0.20$. It is then important to note that the increasing disorder brought by the increasing Si content progressively relaxes the Raman selection rules up to the point that the Ge-Ge band is almost identical between ZZ and YY scattering configurations for the $x = 0.59$ sample. More details about this effect are presented in the following discussion and in sections **SI2** and **SI3**.

Information about Raman selection rules can be obtained by polarization dependent measurements. A full angular dependence of the E$_{2g}$ and A$_{1g}$ / E$_{1g}$ modes' intensities in 2H-Ge NWs is displayed in **SI3**. The selection rules found for all the modes follow the theoretical expectation for hexagonal crystals, which confirms the high crystal quality of our samples.[11,20,21] For $x \leq 0.20$ we can still distinguish the two components of the Ge-Ge band in the YY scattering configuration, but for Si richer compositions the resolution is lost due to the increasing disorder of the system. Namely, in the YY spectra we do not resolve the $E_{2g}$ and $A_{1g}$ / $E_{1g}$ modes anymore, but we observe a consistent redshift of all three bands with respect to ZZ resulting from the enhancement of the $E_{2g}$ phonon modes with respect to the $A_{1g}$/$E_{1g}$ modes. We also performed a full polarization dependence study on a $x = 0.42$ NW: Raman spectra were collected rotating the polarization of the exciting light in 20° steps. The polarization of the collected light was selected either parallel to the NW long axis ($Z$, see Fig. 2) or perpendicular to the NW long axis ($Y$, see Fig. 2). In Figure 4 we show the intensity and the Raman shift of the three main modes as a function of the polarization angle in the two detection configurations. The scattering intensities of the three main spectral components extracted from the parallel detection measurements are highest for polarization parallel to the NW long axis and lowest for perpendicular polarization. Therefore, they follow the selection rules, which instead are partially relaxed for the perpendicular detection data. In Table 1 we report an estimation of the polarization degree of the modes, $R_i$, calculated according to:

$$R_i = \frac{I_{i,max} - I_{i,min}}{I_{i,max} + I_{i,min}} \qquad \text{(Eq. 1.7)}$$

Where $I_{i,max}$ and $I_{i,min}$ are the maximum and minimum intensities of the i-mode respectively, as estimated from the sine squared fit to reduce the error. The polarization degrees of the $A_{1g}$ / $E_{1g}$ modes (detection parallel to Z) are comparable and above 75% (see also Figure 4 A). On the other hand, the $E_{2g}$ modes have a rather small degree of polarization (see also Figure 4 B) that further decreases in vibrations involving Si atoms.

|                | $A_{1g}/E_{1g}$ modes (detection along Z) | $E_{2g}$ modes (detection along Y) |
|----------------|-------------------------------------------|------------------------------------|
| Ge-Ge          | 0.78                                      | 0.40                               |
| Si-Ge          | 0.79                                      | 0.37                               |
| Si-Si          | 0.77                                      | Not available                      |

**Table 1**: polarization ratios of the main modes compared between the two detection configurations. Regarding the Si-Si mode in Y detection, see **SI3** for the reason of its exclusion from this analysis.

This is a combined effect of the alloy disorder and the NW typical aspect ratio. The length of the NW in the Z direction guarantees a degree of translational symmetry such that crystal momentum is a good quantum number and selection rules are respected. Instead, the radial dimension of the NW is smaller than the spot size, the probed phonon **q** vector range is larger than in the parallel detection experiments and selection rules are relaxed. We suggest that vibrations involving Si are affected the most because Si is the minority species in the sample and, therefore, it is the least likely that a Si sublattice is reproduced regularly over a shorter distance. For more detail about the Si-Si vibrations in the perpendicular detection experiments, see **SI3.**

It is important to note that, differently from the behavior of the pure Ge sample, in which the two components ($E_{2g}$ and $A_{1g}/E_{1g}$) under each of the three bands are polarized perpendicularly to each other, in the $x = 0.42$ sample, they are polarized along the same direction. As displayed in Figure 4C and 4D and in greater detail in Figures S7 and S8, the Raman shifts are independent of the excitation polarization angle as well as of detection polarization configuration. This confirms the accuracy of the data analysis and proves that no systematic error has been introduced.

Therefore, we conclude that the alloy related disorder has anisotropic effects on the polarization of the phonon modes and on the relaxation of Raman selection rules, which are more evident for the minority chemical species.

**Spatially dependent Raman maps: homogeneity of the alloy**

In order to assess the homogeneity of the crystal structure and alloy composition of the nanowires, we performed spatially dependent Raman measurements. In Figure 5 we present the results of spatially-resolved Raman measurements performed on $x = 0$ and $x = 0.18$ (see **SI1**) single $Si_xGe_{1-x}$ nanowires as 2D maps: the scattering intensity (false colours) is shown as a function of Raman shift and laser position. Namely, the maps are obtained by acquiring Raman spectra along the NW length with steps of 150 nm ($x = 0$ sample) and 330 nm ($x = 0.18$ sample). In the $x = 0$ sample we find a cubic domain at the "top" of the NW (highlighted by the circles in Figure 5), indicated by the slight blueshift of the phonon frequencies and the enhancement of the $A_{1g}/E_{1g}$ mode in the YY configuration. This mode is allowed only with a low intensity for this configuration in the lonsdaleite phase and is instead the only component present for the diamond phase, at slightly higher frequencies ($F_{2g}$ mode, >300 cm$^{-1}$ [30]).

This observation is in agreement with the structural characterization by TEM studies that exposed cubic Ge (or $Si_xGe_{1-x}$) is due to VLS growth of Ge (or $Si_xGe_{1-x}$) on top of the GaAs core. The $x = 0.18$ sample shows a blueshift of the Ge-Ge band at the upper end of both scans (ZZ and YY). If we had only a change in stoichiometry and not in crystal structure, we would still see an intense $E_{2g}$ contribution in YY at about 287 cm$^{-1}$, whereas we observe one feature at about 297 cm$^{-1}$ with a low energy tail that we assign to the $F_{2g}$ mode of the cubic structure. Therefore, we can discard the hypothesis of the Ge accumulation at the tip of the NW (*i.e.* a segregated phase with different composition than the body of the NW) and confirm the presence of a cubic domain also for $x = 0.18$. The formation of these domains was however already accounted for in the synthesis method of the samples, for which there is still some axial growth during the radial growth of the shell, and this extra portion tends to assume the cubic structure (see also Figure 1A and 1B). For more details about the cubic segments, see Fig S9 in **SI4**. With the exception of this cubic domain at the end of the NW growth, the measured alloy composition and crystal quality of the NWs are homogeneous. The good control of the synthesis conditions in terms of composition and crystalline quality is further corroborated by the results of the quantitative analysis of the spatial dependent Raman studies. In Figure 5E and 5F we plot the peak frequency extracted from analysis of the spectra at selected positions along the NW length: except for the discussed blueshifts, the phonon frequencies stay almost constant throughout the length of the wires. Also, the FWHM of the peaks (see Figure S10 in **SI4**) shows little variation along the NW length. The bands would otherwise broaden or they would shift in energy if at certain positions along the NW the laser illuminated structurally different regions, *i.e.* if the crystal structure or the composition are not homogeneous. We found similar results when performing Raman scans on the $x = 0.59$ sample (not shown).

**Resonant Raman for band-structure probing**

We investigated the electronic band structure of three $Si_xGe_{1-x}$ samples ($x = 0, 0.21, 0.59$) by means of resonant Raman. We performed Raman scattering experiments on single NWs with several excitation wavelengths, ranging from 457 nm to 681 nm. As a reference, a cubic Ge substrate has been used. The cubic Ge sample shows one only maximum at 2.18 eV (see Figure S10 in **SI5**) in agreement with refs 17 and 30. More details on the experimental procedure can be found in **SI5**.

In Figure 6 we report the normalized maximum scattering intensities as functions of the excitation energy. Comparing our data with electronic band structure calculations, we were able to link enhancements in the Raman scattering intensity to electronic transitions at high-symmetry points of the BZ.[13,31] We carried out the assignment of the transitions keeping in mind that the calculations were performed for a system at 0 K and, for $x > 0$, choosing the electronic bands deriving from the most probable atomic arrangements (see refs 13 and 31 for details). To address the possible influence of the As dopants on the resonant Raman profiles, we considered that, as shown in ref 13, the As-related electronic levels merge together in a "dopant" band, which in turn merges with the lowest-lying conduction band.[32] This is consistent with the mono-exponential decay of the PL intensity, meaning the presence of the dopants does not open other emissive channels than just direct band-gap emission. It is reasonable to assume that these considerations would apply not only to the gap, but also to the points of the BZ probed in our work. Despite the band merging might result in transition energy changes with respect to an undoped material, this would be hardly detectable in our experiments, for two reasons: this change could be masked by the configurational fluctuations of the alloys; resonance induced Raman scattering enhancements are generally broad in energy.

The pure 2H-Ge sample exhibits a first resonance around 1.9 eV and another one at 2.33 eV in both the ZZ (Figure 6A) and YY (Figure 6D) polarization configurations, compatible with electronic transitions at the A point and at the M point, respectively.[13,31] The $x = 0.21$ sample shows a resonance at 1.91 eV in both polarizations (Figure 6B and 6E), compatible with a transition at the A point. [13,31] We did not recover the high energy resonance observed for $x = 0$, and unfortunately the lack of laser sources between 2 and 2.2 eV keeps us from clearly identifying other maxima within this energy range. However, the increase of signal at 2.18 eV especially in the YY configuration (Figure 6E) hints at the existence of another maximum in the Raman resonance profile. For the $x = 0.59$ Si sample, we found a low energy resonance around 1.85 eV in YY scattering configuration (Figure 6F), consistent with a transition at the A point, and a high energy resonance at 2.21 eV in ZZ scattering configuration (Figure 6C), in agreement with a transition at the M point.[13,31]

If we now focus on the pure 2H-Ge ($x = 0$) data, it is important to note the slightly stronger enhancement of the $E_{2g}$ phonon mode in YY scattering configuration with respect to the $A_{1g}/E_{1g}$ enhancement in ZZ scattering configuration at 1.91 eV. The scattering intensities measured at 2.33 eV, instead show a stronger coupling of the electronic transitions to the $A_{1g} / E_{1g}$ mode. Based on the electronic band structure calculations we identify these transitions as $\Gamma_{9v}/\Gamma_{7v+} \rightarrow \Gamma_{8c}/\Gamma_{7c}$ at the A point and $\Gamma_{9v} \rightarrow \Gamma_{8c}$ at the M point, respectively, taking into account the degeneracy of the bands at the A point. Considering the growth direction of the NWs and their orientation during the measurements, the different enhancements according to light polarization lead us to believe that the symmetry product $\Gamma_{9v} \otimes \Gamma_{8c}$ is such that the electronic transition at M is favored by light polarized parallel to <0001>, namely the NW long symmetry axis. For the other resonance, there is a considerable enhancement in both polarization configurations with a slight preference for excitation perpendicular to <0001>. This is explainable if we consider that we are detecting the relaxation (*via* the electron-phonon scattering) of carriers not only exactly at the A point, where the bands become degenerate, but also in its vicinities, where they can both express their different symmetry character. The proliferation of possible atomic configurations and, therefore, of their corresponding electronic bands, complicates this analysis for the alloy samples.

In the case of the $Si_xGe_{1-x}$ alloys, it is interesting to look at what specific phonon mode is more sensitive to the variation of excitation energy. For the $x = 0.21$ sample, the Ge-Ge peaks are the most intense in the spectra due to the very small percentage of Si. Accordingly, the Ge-Ge mode exhibits the largest enhancement effect, followed by Si-Ge and finally Si-Si which shows almost no response to the varying excitation energies. The behaviour of the 59% Si sample is not so simple, starting from the change of the dominant peak with respect to light polarization. One can already appreciate this by comparing Figure 3A and 3B, where the ZZ and YY spectra of a 59% Si NW recorded with 514 nm (2.41 eV) of excitation energy are presented. The Si-Si peak is the most intense in ZZ scattering configuration throughout the excitation energy range, while the same is generally true for the Si-Ge peak in YY scattering configuration. The peculiarity is the stronger enhancement of Ge-Ge with respect to Si-Ge found at both resonances (Figure 6C and 6F). We suggest this is due to an outgoing resonance of the Stokes photons after emission of the Ge-Ge phonon energy. It might indeed be that the emission of the Ge-Ge phonon energy brings the Stokes photons in better resonance with the electronic transitions, specifically enhancing this vibration mode.

CONCLUSIONS

We perform Raman scattering measurements on seven GaAs-$Si_xGe_{1-x}$ core-shell nanowire samples of different composition, namely $x = 0$, 0.13, 0.18, 0.21, 0.33, 0.42 and 0.59. In the $x = 0$ sample (pure Ge), we find one main phonon band, composed by the transverse optical $E_{2g}$ mode at ~ 288 cm$^{-1}$ (which represents the signature of the hexagonal phase) and by the degenerate transverse and longitudinal $A_{1g}/E_{1g}$ modes at ~299 cm$^{-1}$. The spectra of $x > 0$ samples are instead formed by three prominent bands, attributed to Ge-Ge, Si-Ge and Si-Si vibrations in order of increasing Raman shift. Each of these bands is composed by an $E_{2g}$-like mode and $A_{1g}/E_{1g}$-like modes, which can be distinguished by polarization-resolved studies. We are able to uncover the frequency dependence on the composition of the two components of each of the three bands, most importantly of the hexagonal signature $E_{2g}$-like modes. Furthermore, as the vibrations of Ge-Ge, Si-Ge, and Si-Si atom pairs happen in different chemical environments, this results in the appearance of additional spectral features, broader and weaker than the three main bands. We assign each feature to the most probable atomic configuration based on qualitative and quantitative considerations about their frequency and intensity as functions of the alloy composition.

We also investigate the Raman selection rules of each phonon mode by carrying out extensive polarization-dependent Raman studies on samples with $x = 0$ and $x = 0.42$. Regarding the former, we found that the $E_{2g}$ mode is polarized perpendicularly to the NW long axis, while the $A_{1g}/E_{1g}$ modes are polarized parallel to it. Regarding the latter instead, we quantitatively estimate how a Si atomic percentage of 42 has the effect of anisotropically relaxing the Raman selection rules, and how this effect is stronger for the minority chemical species (Si-related vibrations). In addition, we also show how, contrarily to the $x = 0$ sample, the $E_{2g}$ and $A_{1g}/E_{1g}$ modes are polarized along the same direction, namely the one at which the polarization of the collected light is fixed.

By performing spatially-resolved measurements on $x = 0$, $x = 0.18$ and $x = 0.59$ samples, we assess the homogeneity and crystalline quality of the NWs. Through the observation of peak blueshift and enhancement of forbidden modes we identify cubic domains at one end of the NWs, which is a consequence of the NW growth process. Otherwise, we illustrate how peak frequencies and linewidths stay constant throughout the NWs, which testifies their very good crystal quality. The results of the resonant Raman scattering experiments performed on 2H-$Si_xGe_{1-x}$ alloys are compared with electronic band structure calculations, and show good agreement between excitation energies coinciding with enhancements of the Raman intensity and possible electronic transitions at various points of the Brillouin zone, [13,31,33] different than the $\Gamma$ point, thus providing a crucial insight in the electronic band structure of this material. Regarding the $x = 0$ sample, we infer the symmetry properties of the electronic bands involved in the transitions by comparing the results obtained in polarization-resolved experiments, specifically observing different coupling of transitions at different points of the BZ with phonons of different symmetry. As far as the $x = 0.21$ and $x = 0.59$ samples are concerned, we observe different resonance profiles with respect to the polarization configuration. For the $x = 0.59$ sample, we observe a sudden enhancement of the Ge-Ge vibrational mode and we argue that it is due to better match of the Stokes photons (after loss of the Ge-Ge energy) with the electronic transition energy.

This work provides a deep investigation of the lattice dynamics and of the electronic resonances of Ge and $Si_xGe_{1-x}$ alloys, which can be grown in the hexagonal phase owing to the nanowires' geometry, and prompts to the use of these nanostructures in optoelectronics and phononic devices.

METHODS

**Synthesis of nanowires:** The core-shell nanowire structures were grown in a Close Coupled Showerhead (CCS) MOVPE reactor *via* catalyst-assisted growth following the VLS mechanism utilizing Au catalyst seeds. The Au catalyst seeds were deposited in nano disks arrays arrangement on a GaAs (111)B substrate *via* the electron beam lithography technique. The growth was performed at a reactor flow of 8.2 slm (standard liters per minute) and a reactor pressure of 50 mbar. For the GaAs NWs, the growth template was annealed at a set thermocouple temperature of 635 °C under an $AsH_3$ flow set to a molar fraction of $\chi_{AsH_3} = 6.1 \times 10^{-3}$. Then, the growth was performed at a set temperature of 650°C with TMGa and $AsH_3$ as material precursors set to molar fractions of $\chi_{TMGa} = 1.9 \times 10^{-5}$, $\chi_{AsH_3} = 4.55 \times 10^{-5}$, respectively, resulting in a V/III ratio of 2.4. After the growth of the GaAs core NWs, they were chemically treated with a cyanide based solution to remove the Au catalyst particles to avoid gold contamination in the $Si_xGe_{1-x}$ shells. Eventually, the GaAs NW core was used as a hexagonal material template and was overgrown with a $Si_xGe_{1-x}$ shell by introducing the suitable gas precursors for the shell growth which are $GeH_4$ and $Si_2H_6$. The $Si_xGe_{1-x}$ shell was grown at a set temperature of 650 °C at a molar fraction of $\chi_{SiGe} = 1.55 \times 10^{-4}$ for a certain time according to the desired thickness.

**Structural characterization:** For the cross-section TEM studies, NWs were prepared using Focused Ion Beam (FIB). In the different imaging modes, HR-TEM and STEM, analyses were conducted using a JEM ARM200F probe-corrected TEM operated at 200 kV. For the chemical analysis, EDX measurements were carried out using the same microscope equipped with a 100 mm$^2$ EDX silicon drift detector. TEM lamellae were prepared in a FEI Nova Nanolab 600i Dual beam system. For this, the NWs were initially transferred with the aid of a nano-manipulator from the growth substrate to a piece of Si and then arranged to lie parallel to each other. These NWs were covered with electron- and ion-beam induced metal deposition to protect them during the FIB milling procedure. The lamella was cut out by milling with 30 kV Ga ions and thinned down with subsequent steps of 30, 16, and 5 kV ion milling in order to minimize the Ga-induced damage in the regions imaged with TEM.

**Computational details:** We performed density-functional calculations with the Abinit code,[34,35] using a plane wave cutoff of 38 Ha, the local density approximation (LDA) for the exchange-correlation energy, and norm-conserving pseudopotentials.[36] The Raman susceptibility tensor was computed within density-functional perturbation theory (DFPT)[37] from the third derivative of the total energy, twice with respect to the application of an electric field (*i.e.* incident and scattered light polarization vectors) and once with respect to the phonon displacement coordinates, making use of the 2n+1 theorem.[38] The off-resonance Raman intensity of each mode was calculated as

$$I_n \propto |\varepsilon_i \, R_n \, \varepsilon_s|^2$$

where $\varepsilon_i$ and $\varepsilon_s$ are the polarization vectors of the incident and scattered light, respectively, and $R_n$ is the Raman susceptibility tensor. The spectra were then generated by summing lorentzian functions centered at each computed frequency, with a full width at half maximum of 3 cm$^{-1}$. For a discussion on the choice of the **k**-point mesh see ref. 18.

We used a 2x2x1 supercell of the 4-atom unit cell of 2H crystals, randomly populating the lattice sites in order to reproduce the target composition of the alloy. In particular we used 1 Si atom and 15 Ge atoms for $x$=0.0625, 3 Si atom and 13 Ge atoms for $x$=0.19, and 8 Si atom and 8 Ge atoms for $x$=0.50. In the case of $x$=0.19 and $x$=0.50 we independently generated 5 and 6 random configurations, respectively, and performed a weighted average of the corresponding polarized Raman spectra taking as weights the cohesive energies of each alloy. In this way configurations that are energetically favored, and thus more likely to occur, contributed more to the averaged spectrum. No averaging was done for $x$=0.0625 where one single Si atom is added substitutionally to the 2H-Ge lattice. We compared our results with those obtained averaging on more configurations of a smaller unit cell, finding qualitatively similar results, but somewhat more noisy spectra. For this reason, we based our analysis on the larger 16-atom cells. The calculations were performed for a system at 0 K.

**Raman spectroscopy details:** Raman experiments were performed by exciting the samples with different lasers: the 633 nm line of a HeNe laser, the different wavelengths of Ar and Ar$^-$Kr lasers, the 532 nm of a diode-pumped solid state laser, the 561 nm of an optically pumped semiconductor laser, and the 681 nm wavelength of a wavelength-stabilized diode laser. An incident power of 40 μW was used to avoid heating or damaging effects. The samples were illuminated with a high numerical aperture (0.95) 100x objective, which allows a spatial resolution of about 600 nm for the 514.5 nm wavelength. The backscattered light was collected by a T64000 triple spectrometer in subtractive mode, equipped with 1800 g/mm gratings and a liquid nitrogen-cooled CCD detector. The polarization of the incident laser beam and of the backscattered light was controlled and selected by means of lambda-half waveplates and linear polarizers.

**Supporting Information**

The Supporting Information is available free of charge at https://pubs.acs.org/doi/10.1021/acsnano.*******. It contains further structural characterization of the nanowires, details on computational and spectroscopic methods and analysis, and Raman spectra taken on different nanowires, with different scattering geometries, and with different excitation wavelengths (PDF).

AUTHOR INFORMATION

**Corresponding Author**

*E-mail: ilaria.zardo@unibas.ch

**Author Contributions**

IZ conceived the experiment. DDM and MDL performed the Raman measurements, which were analyzed by DDM. MLS and RR performed the theoretical calculations. EMTF and EPAMB grew the samples. MAV performed TEM investigations. DDM, MDL, and IZ wrote the manuscript with contributions of all authors. All authors have given approval to the final version of the manuscript.

ACKNOWLEDGMENT

This project has received funding from the European Research Council (ERC) under the European Union's Horizon 2020 research and innovation program (grant agreement No 756365). M. D. L. acknowledges support from the Swiss National Science Foundation Ambizione grant (Grant No. PZ00P2_179801). R.R. acknowledges financial support by the Ministerio de Economía, Industria y Competitividad (MINECO) under grant FEDER-MAT2017-90024-P and the Severo Ochoa Centres of Excellence Program under grant SEV-2015-0496 and by the Generalitat de Catalunya under grant no. 2017 SGR 1506. E. P. A. M. B. and E. M. T. F. acknowledge European Union's Horizon 2020 research and innovation program under grant agreement No 735008 (SiLAS). E. P. A. M. B. and M. A. V. acknowledge Solliance, a solar energy RD initiative of ECN, TNO, Holst, TU/e, IMEC, Forschungszentrum Jülich, and the Dutch province of Noord-Brabant for funding the TEM facility. R.R. thanks Silvana Botti for useful discussions.

FIGURES

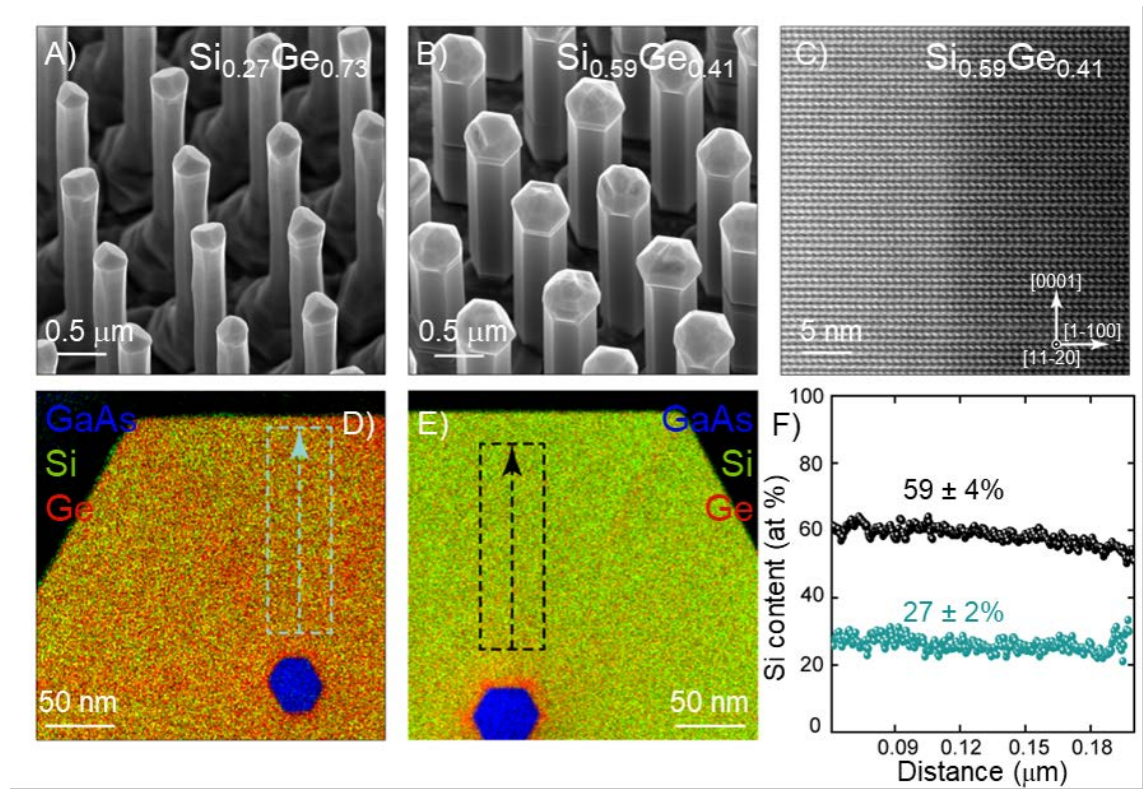

**Figure 1.** Epitaxial 2H-GaAs-Si$_x$Ge$_{1-x}$ core-shell nanowire structures: (A-B) 30 degree tilted-view SEM images of arrays of 2H-Si$_{0.27}$Ge$_{0.73}$ (A) and Si$_{0.59}$Ge$_{0.41}$ (B) shells grown around thin (40 nm) WZ-GaAs nanowire cores on a GaAs (111)B substrate in the [0001] crystallographic direction. C) A HAADF-STEM image of the interface of the core-shell structures in B) acquired in the [11$\bar{2}$0] zone axis, displaying the ABAB stacking along the [0001] direction confirming the hexagonal crystal structure. (D-E) EDX elemental maps of cross sectional lamellas of representative GaAs-Si$_{0.27}$Ge$_{0.73}$ and GaAs-Si$_{0.59}$Ge$_{0.41}$ core-shell NWs showing the GaAs NW core in blue, Si atoms in green and Ge atoms in red. F) Radial line scans along the arrows marked by the dashed boxes in the EDX maps shown in D) and E) quantifying the silicon content in both samples.

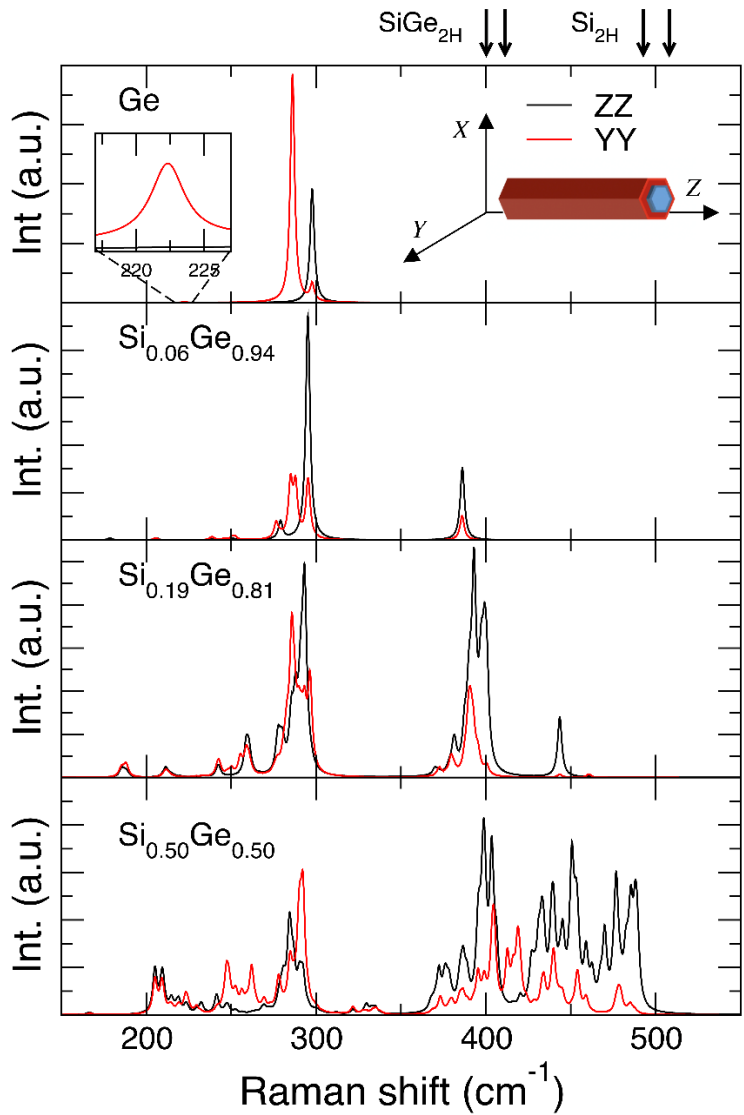

**Figure 2.** Computed DFPT Raman spectra in ZZ (black lines) and YY (red lines) scattering geometry of (from top to bottom) 2H-pure Ge, $Si_{0.06}Ge_{0.94}$, $Si_{0.19}Ge_{0.81}$, and $Si_{0.5}Ge_{0.5}$ alloys. The small peak at 222.3 cm$^{-1}$ in the spectra of pure Ge YY (see the zoomed inset in first panel) is due to the back-folding of the LA branch. The cartoon in the first panel shows the orientation of the axes with respect to the NW samples.

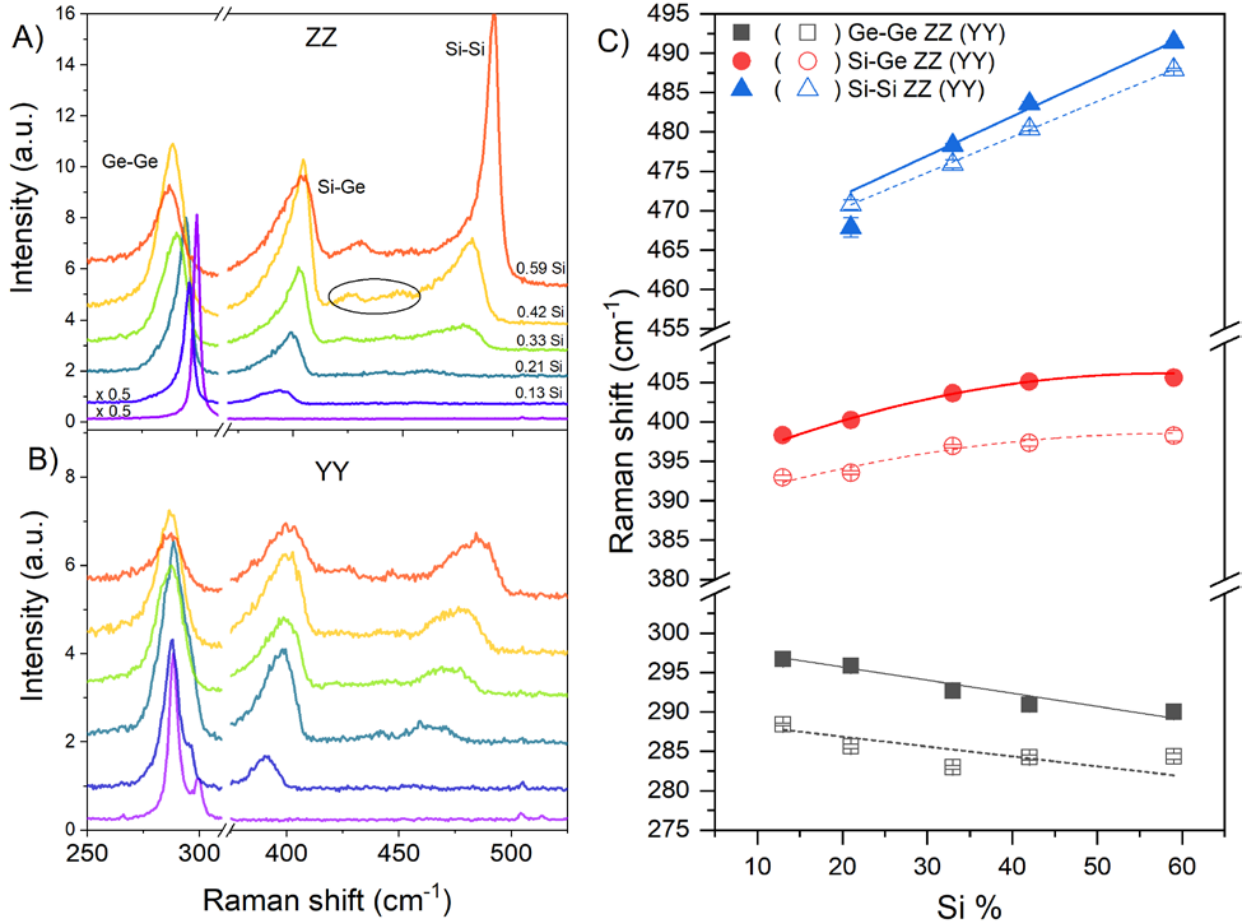

**Figure 3.** Comparison of the Raman spectra of five $Si_xGe_{1-x}$ alloy nanowires recorded in A) ZZ scattering configuration and B) YY scattering configuration. From bottom to top $x = 0.13$; $0.21$; $0.33$; $0.42$; $0.59$. Both panels also show the spectrum of a pure 2H-Ge NW acquired in the corresponding scattering configuration (purple line, bottom). The main bands are labeled with their respective oscillator. C) Frequency of the main bands as a function of composition (symbols) and their fits (lines) extracted from the spectra shown in A) and B). The full symbols and solid lines represent the $A_{1g}$ / $E_{1g}$ components of the bands (spectra recorded in the ZZ configuration). The empty symbols and dashed lines represent the $E_{2g}$ components (spectra recorded in the YY configuration). Error bars are taken from the fit results.

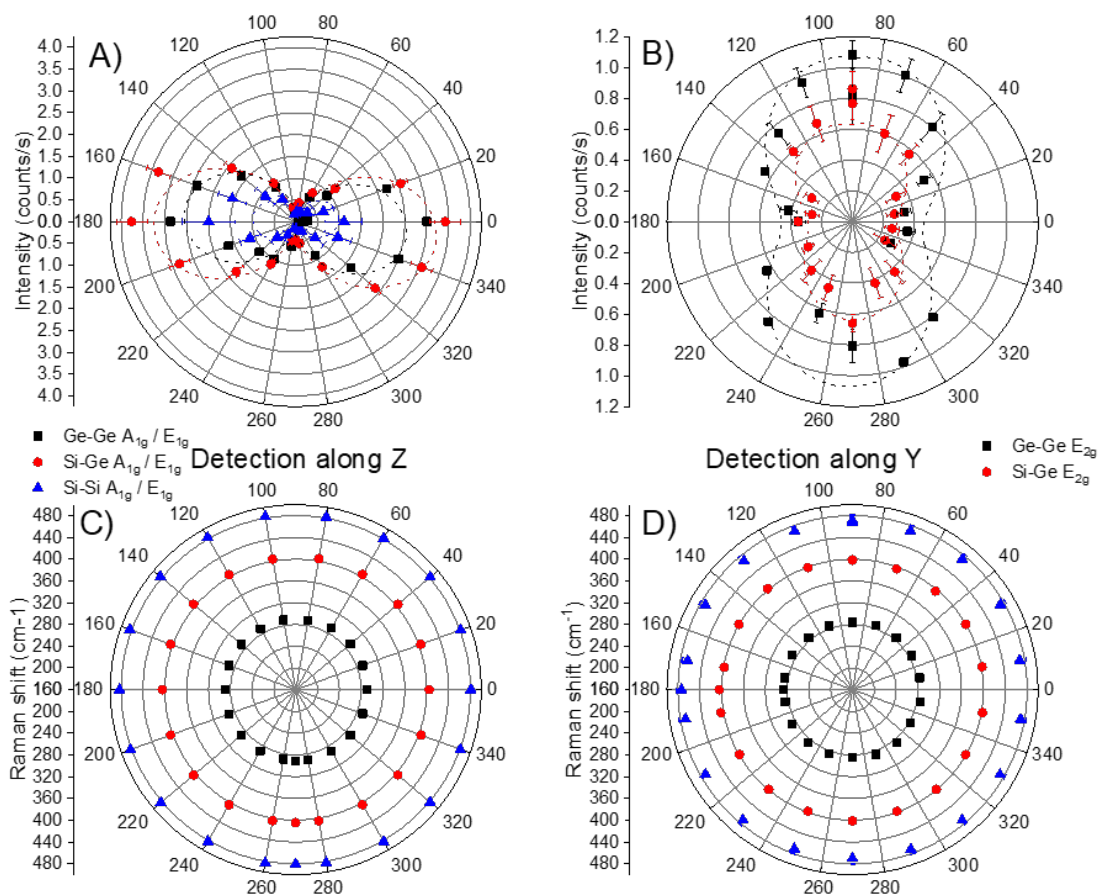

**Figure 4.** Polarization dependence of the intensity of the main modes of the x=0.42 sample with $A_{1g}/E_{1g}$ (A) and $E_{2g}$ (B) symmetry collected with polarization of scattered light along Z and Y, respectively. The dashed lines are sine squared fits of the experimental data. C) and D) Raman shifts of the $A_{1g}/E_{1g}$ and $E_{2g}$ modes with the polarization of the collected light parallel and perpendicular to the NW long axis, respectively.

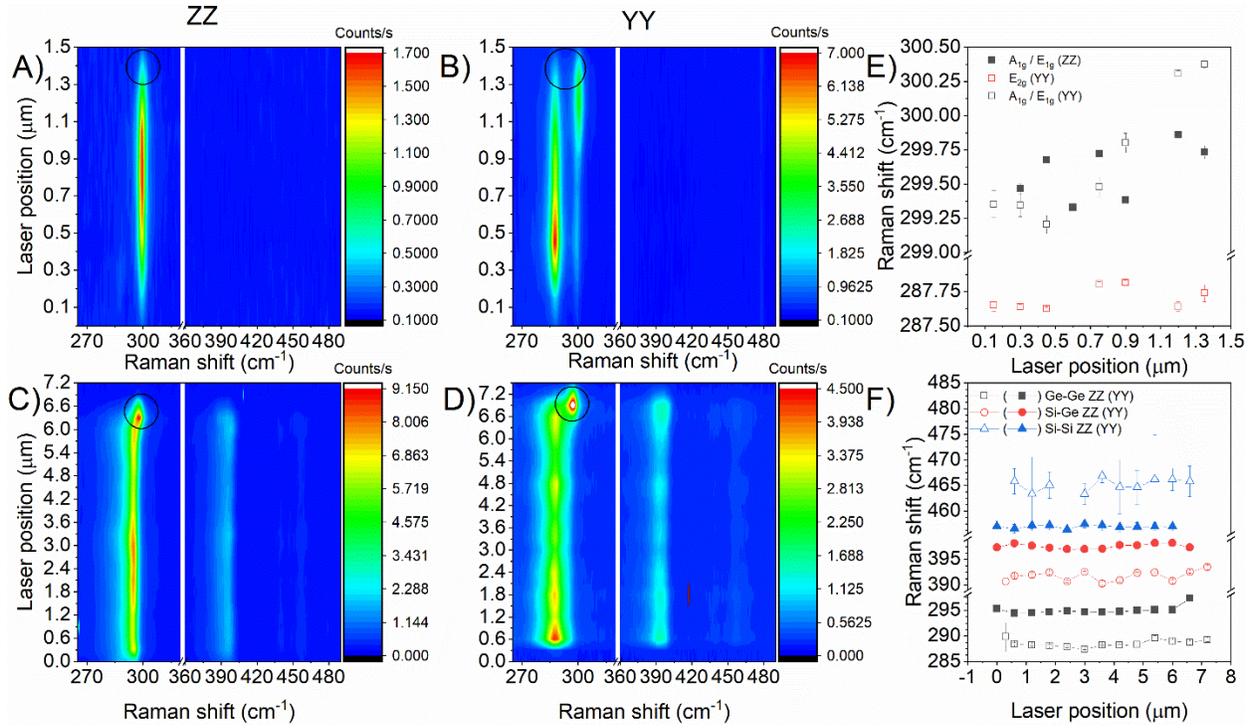

**Figure 5.** A) to D): Spatially-resolved Raman measurements performed on A), B) a $x = 0$ NW (pure Ge), and on C), D) a $x = 0.18$ NW in the polarization configurations indicated above. E), F): peak positions extracted from fits at selected positions.

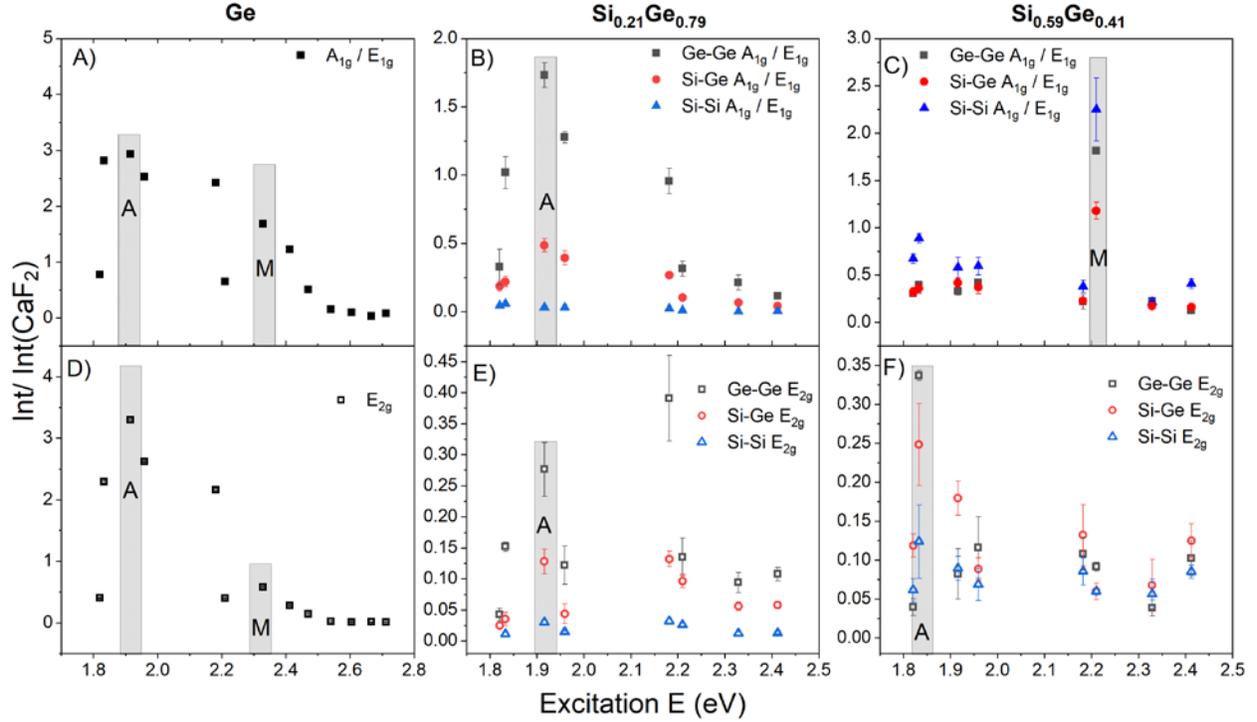

**Figure 6.** Maximum scattering intensity of the main bands as a function of excitation energy for three $Si_xGe_{1-x}$ NWs. Top row: data from spectra recorded in the ZZ configuration; Bottom row: data from spectra recorded in the YY configuration. The grey areas highlight resonance-related enhancements of the intensity and the letters indicate points of the Brillouin zone at which the corresponding electronic transitions happen.[13,31,33] A), D) $x = 0$: the resonance profile is peaked at 1.91 eV and 2.33 eV. These resonances can correspond to electronic transitions at the A and M point respectively; B), E) $x = 0.21$: there is a resonance at 1.91 eV, compatible with an electronic transition at the A point; C), F) $x = 0.59$: the resonance profile is peaked at 1.83 eV in YY and at 2.21 eV in ZZ. These maxima agree with electronic transitions at the A and M point respectively.